\documentclass[pra,twocolumn,showpacs,preprintnumbers,amsmath,amssymb]{revtex4}
\usepackage{graphicx}
\usepackage{dcolumn}
\usepackage{bm}
\usepackage{latexsym,epsfig}

\begin{document}
\preprint{preprint-Feb 04, 2008}
\title{Phase diagram of a  bosonic ladder with two coupled chains.}
\author{Meetu Sethi Luthra}
\email{sethi.meetu@gmail.com} \altaffiliation [permanent address
]{Bhaskaracharya College of Applied Sciences, Phase-I,
Sector-2,Dwarka,Delhi,110075, India.}
 \affiliation{ Indian Institute of
Astrophysics, II Block, Kormangala, Bangalore, 560 034, India.}
\author{Tapan Mishra}
\email{tapan@iiap.res.in} \affiliation{ Indian Institute of
Astrophysics, II Block, Kormangala, Bangalore, 560 034, India.}
\author{Ramesh V. Pai}
\email{rvpai@unigoa.ac.in} \affiliation{ Department of Physics, Goa
University, Taleigao Plateau, Goa 403 206, India. }
\author{B. P. Das}
\email{das@iiap.res.in} \affiliation{Indian Institute of
Astrophysics, II Block, Kormangala, Bangalore, 560 034, India.}
\date{\today}

\begin{abstract}
We study a bosonic ladder with two coupled chains using the finite
size density matrix renormalisation group method. We show that in a
commensurate bosonic ladder the critical on-site interaction ($U_C$)
for the superfluid to Mott insulator transition becomes larger as
the inter-chain hopping ($t_\bot$)increases. We analyze this quantum
phase transition and obtain the phase diagram in the $t_\bot -U$
plane.

\end{abstract}
\pacs{03.75.Nt, 05.10.Cc, 05.30.Jp,73.43Nq}

\keywords{Suggested keywords}

\maketitle
\section{introduction}
Quantum phase transitions in ultracold atoms provide important
insights into the behaviour of matter at very low
temperatures~\cite{bloch,sondhi,sachdev,lewenstein}. An important
example of this class of transitions is the transition from a
superfluid(SF) to a Mott insulator(MI) which has been observed in
cold bosonic atoms in $3D$ optical lattices~\cite{greiner,stoferle}
as predicted by Jaksch \textit{et al}~\cite{jaksch}. Subsequently,
this transition has been observed in a $1D$ optical
lattice~\cite{stoferle}. Detailed theoretical studies of this
transition have been carried out by Pai \textit{et al}~\cite{pai,pandit}.
An important question to address is how do the characteristics of
this transition alter in going from one to two dimensions. In
coupled bosonic chains, competition exists between the ratio of
atomic interactions to the intra-chain hopping and the inter-chain
hopping. A large value of the former favors a Mott insulator state
overcoming the effect of the inter-chain hopping while if the latter
dominates it would tend to delocalize the bosons and drive the
system to a superfluid state. It is not practical to perform
numerical studies for the above mentioned transitions in a very
large number of coupled chains, so one must restrict to a finite
number of such chains. The aim of the present work is to study the
effect of the inter-chain hopping on the SF-MI transition for a
bosonic ladder consisting of two coupled chains. Although a
substantial amount of theoretical and numerical work has been done
in this direction for the case of spin-less fermionic ladders and
spin ladders \cite{donohue,Carr,park},no work has been done to our
knowledge for the bosonic ladders except a recent work using the
Bosonization method. \cite{giamarchi,orignac}.
\begin{figure}[htbp]
  \centering
  \epsfig{file=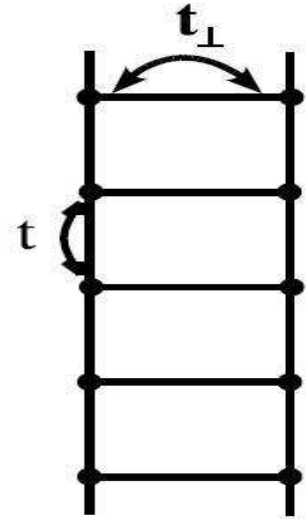,width=6cm,height=8cm}
  \caption{Schematic picture of a two-leg bosonic ladder. $t$ and $t_\bot$
  are, respectively, inter-chain and intra-chain hopping amplitudes.}
  \label{fig:fig1}
\end{figure}
The Hamiltonian of the bosonic ladder(as shown in
Fig.\ref{fig:fig1})is given by
\begin{eqnarray}\label{eq:ham}
\nonumber
\cal{H} &=&-t\sum_{i,\alpha}(a_{i,\alpha}^{\dagger}a_{i+1,\alpha}+h.c)\\
\nonumber && + \frac{U}{2}\sum_{i,\alpha}n_{i,\alpha}(n_{i,\alpha}
-1)\\ &&-t_\bot \sum_{i} (a_{i,1}^{\dagger}a_{i,2}+h.c).
\end{eqnarray}
In this model (\ref{eq:ham}) $a_{i,\alpha}^{\dagger}$
($a_{i,\alpha}$) represents bosonic creation (annihilation) operator
for the site $i$ of the chain with index ${\alpha =1,2}$. $t$ and
$U$ are the intra-chain hopping amplitude between the nearest
neighboring sites of chain $\alpha$ and the on-site interaction
between the bosons respectively. The last term in this model
(\ref{eq:ham}) represents inter-chain hopping with an amplitude
$t_\bot$ between corresponding sites on the two chains. We set our
energy scale by taking $t=1$.

This model has  been studied using the Bosonization
technique~\cite{giamarchi, orignac} at or close to commensurate
filling of one boson per site. This study predicts a transition from
a Mott insulator to a superfluid phase when the inter-chain hopping
is increased and it is in the Beresenskii-Kosterlitz-Thouless (BKT)
universality class at commensurate filling. In the present work we
verify these predictions and thus complement the earlier analytical
results. For this purpose, we study the variation of critical
on-site interaction $U_C$ for the superfluid phase to the Mott
insulator phase transition with the change in the inter-chain
hopping amplitude $t_\bot$ using the finite size density matrix
renormalization group method (FSDMRG)
\cite{pandit,white,dmrgreview}and obtain the phase diagram in the
$(t_\bot-U)$ plane. To the best of our knowledge, the present work
is the first application of FSDMRG method to bosonic ladders.

The  remaining part of the paper is organized in the following
manner.The FSDMRG method in the context of bosonic ladders is
briefly described in section \ref{sec:fsdmrg}. Our results are
described and discussed in Section \ref{sec:results} and our
conclusions are stated  in Section \ref{conclusion}.

\section {FSDMRG Method}
\label{sec:fsdmrg}
 We use the FSDMRG technique to obtain
the energies and the correlation functions of the ground state. This
method is very efficient and has proven to give accurate results for
1D quantum lattice systems and has been applied to low-dimensional
strongly correlated fermionic and bosonic
systems~\cite{pandit,tezuka}. We give below some pertinent details
of this method adapted to the system of two coupled chains that we
have considered.

\begin{figure}[htbp]
  \centering
   \epsfig{file=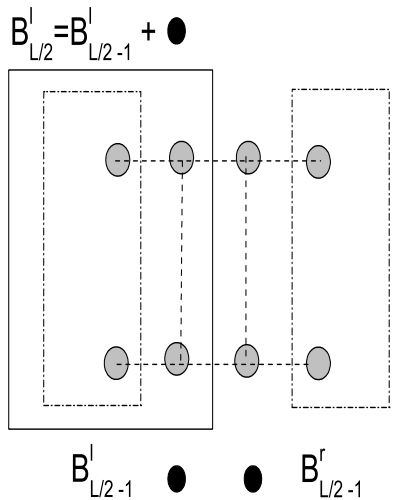,width=8cm,height=8cm}
    \caption{A scheme of superblock configuration for the FSDMRG algorithm}
     \label{fig:fig2}
\end{figure}

We begin with a super-block configuration $B_{\frac{L}{2}-1}^{l} ~~
\bullet ~~ \bullet ~~ B_{\frac{L}{2}-1}^{r}$ of $L$ rungs as shown in Fig.\ref{fig:fig2}.
The left $B_{\frac{L}{2}-1}^{l}$
and the right
block $B_{\frac{L}{2}-1}^{r}$ have $\frac{L}{2}-1$ rungs each
and the $\bullet$ represents one rung of two sites,
one from each chain. Thus in every iteration, the new left and right
blocks are $B_{\frac{L}{2}}^{l}=B_{\frac{L}{2}-1}^{l} ~ \bullet $
and $B_{\frac{L}{2}}^{r}=\bullet ~ B_{\frac{L}{2}-1}^{r}$
respectively. We increase the size of the system in every iteration
by adding two rungs which increases the number of lattice sites by
$4$. To keep the density $\rho=1$ fixed, we also increase the number
of bosons in the system by $4$. The truncation of states of left
(right) block in each iterations corresponds to choosing $M$ highest
weighted states out of $2~\times~n_{\mbox{max}}~\times~ M$ of the
left (right) density matrix. Here $n_{\mbox{max}}$ is the number of
states kept at each site, which is in general infinity, but we
truncate it for a feasible numerical calculation. We keep
$n_{\mbox{max}}=4$ in this calculation which is found to be
sufficient for the values of $U$ considered here~\cite{pandit}. The
value of $M$ is chosen such that the truncation error in our
calculation is always less than $10^{-5}$.

\section{Results and Discussion}
\label{sec:results} The Mott insulator phase has a finite gap in its
energy spectrum. The single particle gap is defined as
\begin{equation}
G_L=E_L(N+1)-E_L(N) -(E_L(N) - E_L(N-1)) \label{eq:gap}
\end{equation}
where $E_L(N)$ is the ground-state energy of two leg bosonic ladder
with length $L$ having $N$ bosons. The MI phase is signaled by the
opening up of the gap $G_{L\to\infty}$. However, $G_{L}$ is finite
for finite systems and we must extrapolate to the $L\rightarrow
\infty$ limit, which is best done by using finite-size
scaling~\cite{pandit}. In the critical region, i.e., SF region, the
gap
\begin{equation}
  \label{eq:scaling-corr}
  G_L \approx L^{-1}f(L/\xi),
\end{equation}
where the scaling function $f(x) \sim x , \, x \to 0$ and $\xi$ is
the correlation length. $\xi \rightarrow \infty$ in the SF region.
Thus plots of $LG_L$ versus $U$, for different system sizes $L$,
consist of curves that intersect at the critical point at which the
correlation length for $L=\infty$ diverges and gap $G_\infty$
vanishes. The phase diagram, as discussed below, is obtained from
such plots.

\begin{figure}[htbp]
  \centering
   \epsfig{file=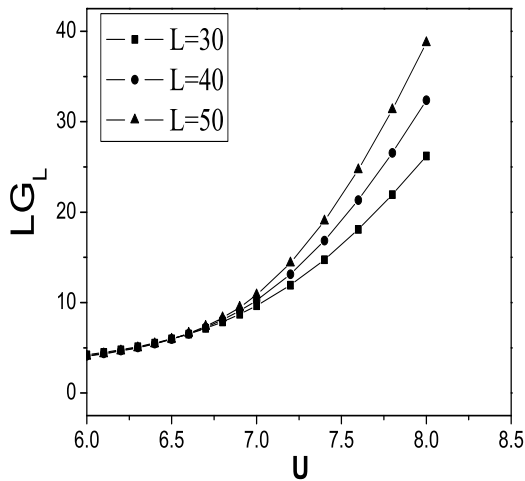,width=8cm,height=8cm}
   \caption{Scaling of gap $LG_L$ as a function of $U$ for $t_\bot= 0.4$ and different lengths.
   The coalescence of curves for different lengths for $U < U_C \sim 6.6$ shows a superfluid
   phase and
    a Mott insulator with finite gap for $U > U_C$.}
    \label{fig:fig3}
\end{figure}

\begin{figure}[htbp]
  \centering
   \epsfig{file= 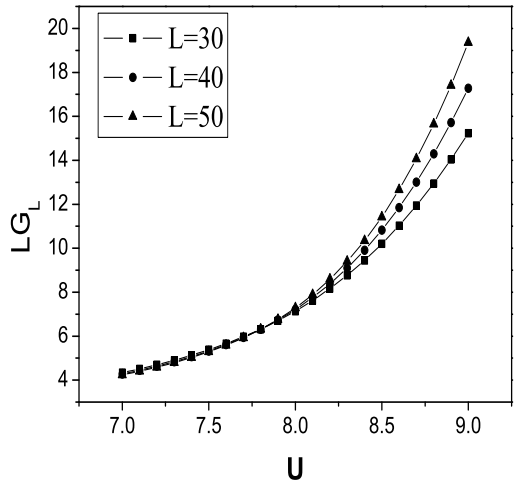,width=8cm,height=8cm}
   \caption{Scaling of gap $LG_L$ as a function of $U$ for $t_\bot = t= 1.0$ and different lengths.
   The coalescence of curves for different lengths for $U < U_C \sim 7.9$ shows a superfluid
   phase and
    a Mott insulator with finite gap for $U > U_C$. Comparing this figure with Fig.\ref{fig:fig3}, we observe that
    the critical $U_C$ increases with $t_\bot$.}
   \label{fig:fig4}
\end{figure}

It is now well known that the single chain Bose-Hubbard model with
density $\rho=1$ shows a SF-MI transition with the critical
on-site interaction $U_C \sim 3.4$\cite{pandit}. In order to
understand the effect of the inter-chain hopping on this transition,
we varied $t_\bot$ from $0$ to $20$ and obtained the corresponding
critical on-site interaction $U_C(t_\bot)$ for the SF-MI
transition. We found that  $U_C$ increases with $t_\bot$ and
saturates in the limit $t_\bot\rightarrow \infty$. These results are
highlighted by the plots of scaling of gap $LG_L$ versus $U$ for
different values of $t_\bot$ and lengths $L$. For example, in the
Fig.\ref{fig:fig3} we plot $LG_L$ versus $U$ for $t_\bot=0.4$. The
coalescence of $LG_L$ curves for different values of $L$ below $U <
6.6$ demonstrates the SF-MI transition with $U_C(t_\bot=0.4) \sim
6.6$ which is much larger than the corresponding value for the
single chain $U_C(t_\bot=0) \sim 3.4$.  Fig.\ref{fig:fig4} represents
similar plots for $t_\bot=1$. For this case the critical on-site interaction
increases further to $U_C((t_\bot=1)\sim 7.9$.

\begin{figure}[htbp]
  \centering
   \epsfig{file= 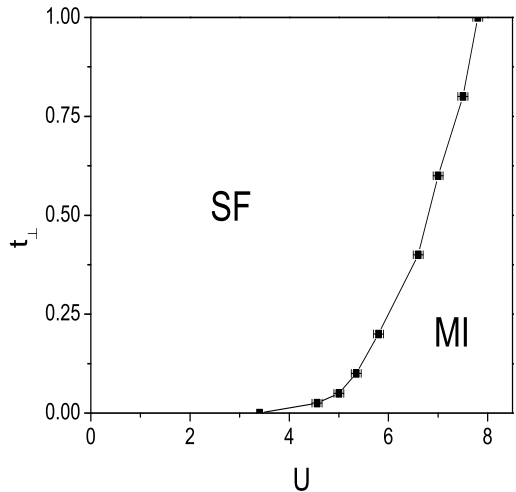,width=8cm,height=8cm}
   \caption{Phase diagram of model (\ref{eq:ham}) as a function of inter-chain hopping $t_\bot$ and on-site interaction
   $U$ for density $\rho=1$. Note that we have set intra-chain hopping $t=1$.}
   \label{fig:fig5}
\end{figure}

\begin{figure}[htbp]
  \centering
   \epsfig{file= 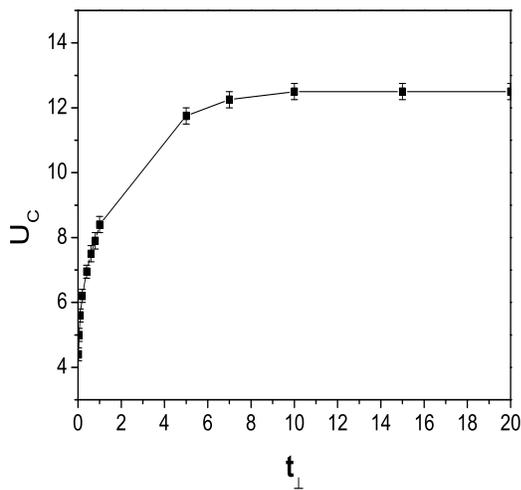,width=8cm,height=8cm}
   \caption{Variation of critical on-site interaction $U_C$ with respect to inter-chain hopping $t_\bot$.
   $U_C$ increases sharply for small values of $t_\bot$ and saturate to $12.5\pm0.3$ as $t_\bot \rightarrow \infty$.}
   \label{fig:fig6}
\end{figure}

\begin{figure}[htbp]
  \centering
   \epsfig{file= 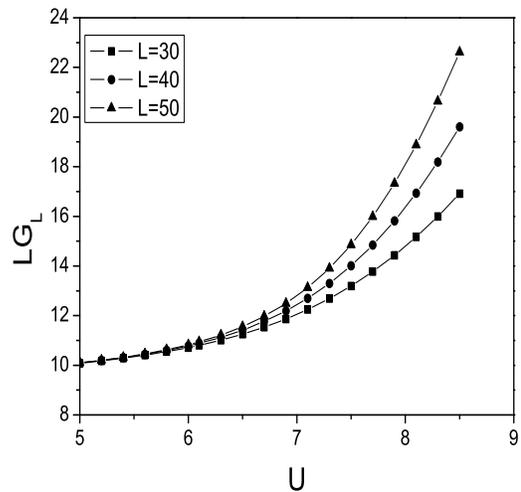,width=8cm,height=8cm}
   \caption{Scaling of gap $LG_L$ as a function of $U$ for the single chain Bose-Hubbard model with density $\rho=2$.
   The coalescence of curves for different lengths for $U < U_C \sim 6.3$ shows a superfluid
   to a Mott insulator transition.}
   \label{fig:fig7}
\end{figure}

From similar plots of $LG_L$ versus $U$, we obtain the phase diagram
for model (\ref{eq:ham}) in the $t_\bot-U$ plane and it can be seen
in Fig.\ref{fig:fig5}. $U_C$ for the SF-MI transition initially
increases sharply as the inter-chain hopping $t_\bot$ increases.
This phase diagram verifies the prediction of MI-SF transition with
respect to increase in the inter-chain hopping
$t_\bot$\cite{donohue}. For higher values of $t_\bot$, $U_C$ tends
to saturate. For $t_\bot \gg t,U$, each rung has two one particle
states: corresponding to bonding or anti-bonding. As predicted in
the Bosonization~\cite{donohue} study, this problem then maps onto a
single chain Bose Hubbard model with commensurate density $\rho=2$
and on-site interaction $U/2$. To confirm this prediction we plot
the variation of $U_C$ with respect to $t_\bot$ in
Fig.\ref{fig:fig6}. Critical on-site interaction $U_C$ for large
$t_\bot$ converges to a value equal to $12.5\pm 0.3$. Plotting
$LG_L$ versus $U$ for single chain Bose-Hubbard model for $\rho=2$
in Fig.\ref{fig:fig7} we find that $U_C \sim 6.3$, which is one half
the converged value of $U_C(t_\bot=\infty) \sim 12.5$ for the
bosonic ladder confirming the prediction made in Bosonization
study\cite{donohue}.

The Mott insulator to superfluid transition is found to be
Beresinskii-Kosterlitz-Thouless (KT) universality class at
commensurate filling\cite{donohue}. The correlation function that
characterizes the superfluid phase is given by
$\Gamma_\alpha(r)=\langle a^\dagger_{i,\alpha}a_{i+r,\alpha}\rangle$
which decays as a power law in the limit $r\rightarrow \infty$. Here the
expectation value is taken with respect to the ground state.
However, in the Mott insulator phase it has an exponential decay due
to the finite gap in the energy spectrum. The power law decay of
this correlation function has been obtained using the Bosonization
method~\cite{donohue} and it is predicted to go as
\begin{equation}\label{eq:corr}
     \Gamma_\alpha(r) \propto \frac{1}{r^{1/4K_s}}
\end{equation}
with the Luttinger Liquid parameter $K_s$  stated to be $1$ at the
superfluid to Mott insulator transition point.

\begin{figure}[htbp]
  \centering
   \epsfig{file= 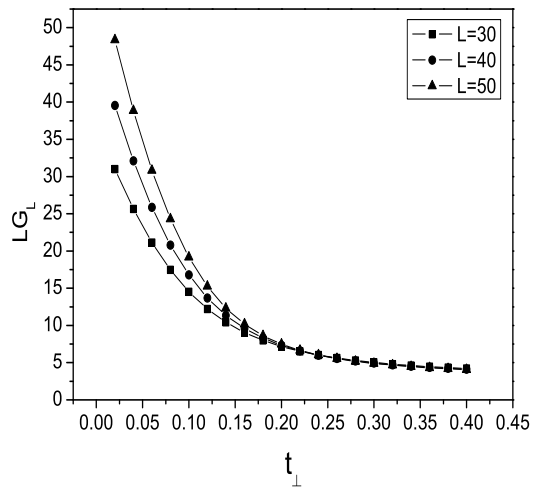,width=8cm,height=8cm}
   \caption{Scaling of $LG_L$ as a function of $t_\bot$ for $U=6$.}
   \label{fig:fig8}
\end{figure}

\begin{figure}[htbp]
  \centering
   \epsfig{file= 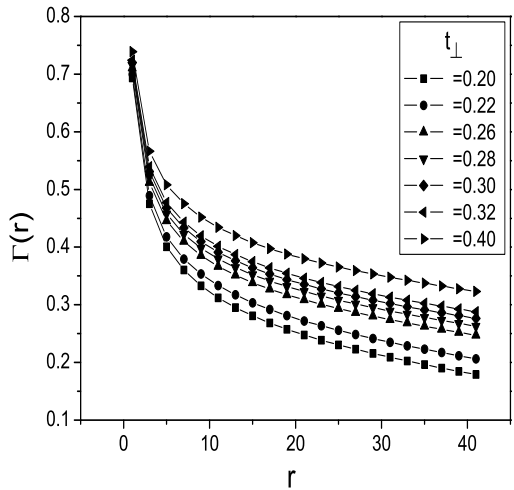,width=8cm,height=8cm}
   \caption{Power law decay of  $\Gamma_\alpha(r)$ for various values of $t_\bot$ for $U = 6$.}

   \label{fig:fig9}
\end{figure}

\begin{figure}[htbp]
  \centering
   \epsfig{file= 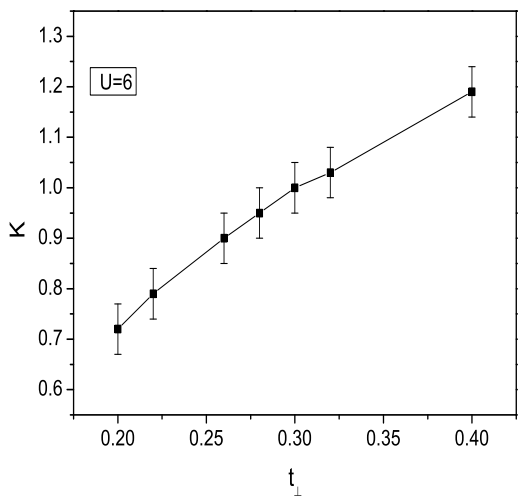,width=8cm,height=8cm}
   \caption{Variation of  Luttinger liquid parameter $K_s$ as a function of  $t_\bot$ near MI to SF transition for $U=6$.}
   \label{fig:fig10}
\end{figure}

In order to obtain the Luttinger Liquid parameter $K_s$ we fix $U=6$
and vary the inter-chain hopping $t_\bot$ and obtain the Mott
insulator to superfluid transition. The scaling of gap $LG_L$ as a
function of $t_\bot$ is given in Fig.\ref{fig:fig8}. The critical
inter-chain hopping $t^C_\bot =0.24\pm0.05$ for the MI to SF
transition. The correlation functions $\Gamma_\alpha(r)$ for $U=6$
different values of $t_\bot$ are given in
Fig.\ref{fig:fig9}. The Luttinger Liquid parameter $K_s$ which is
obtained by fitting $\Gamma_\alpha(r)$ with the expression given in
Eq.\ref{eq:corr} is plotted as a function of $t \bot$ in Fig.\ref{fig:fig10}. From these values
the critical $t^C_\bot$ for which $K_s=1$ is given by $0.3\pm0.03$
which is consistent with values obtained from the scaling of the gap.

The results we have obtained could have experimental
implications. It is now possible to prepare bosonic ladders by
growing optical superlattices in the form of double well potential
along one direction\cite{rey}. The tunneling between the double well
potential will control the inter-chain hopping. By changing
dynamically the optical lattice parameters one can control all the
interaction and hopping parameters of the model (\ref{eq:ham}).

\section{Conclusions}
\label{conclusion} We have studied the ground state properties of a
two chain bosonic ladder with commensurate filling of one boson per
site using the finite size density matrix renormalization group
method. The critical on-site interaction for SF-MI phase transition
increases sharply for small values of inter-chain hoping amplitude
$t_\bot$. However, it saturates in the limit $t_\bot \rightarrow
\infty$. Thus in the presence of large inter-chain hopping, the
system continues to be in the superfluid state even though the
single chain is a Mott insulator. Thus we confirm the prediction of
Mott insulator to superfluid transition as a function of $t_\bot$.
We have obtained the Luttinger Liquid parameter and compared it with
the analytical results. In addition to verifying and complementing
the predictions made by the Bosonization technique, we have pointed
out the possible experimental verification of our results. We hope
that our analysis of the SF-MI transition in bosonic ladders will
stimulate experimental studies in this direction.

\section{acknowledgments} One of us (ML) thanks the Indian Institute of Astrophysics,
Bangalore where this work was done during her visit. This work was
supported by DST, India (Grants No. SR/S2/CMP-0014/2007).

\end {document}